\documentclass[runningheads]{llncs}
\usepackage[T1]{fontenc}
\usepackage[ruled,vlined]{algorithm2e}
\usepackage{comment}
\usepackage{graphicx}
\usepackage{subcaption}
\usepackage{subfig}

\begin{document}
\title{A Bio-Inspired Leader-based Energy Management System for Drone Fleets}
%
%
\author{Rosario Napoli\inst{1}\orcidID{0009-0006-2760-9889} \and Antonio Celesti \inst{1}\orcidID{0000-0001-9003-6194} \and Massimo Villari \inst{1}\orcidID{0000-0001-9457-0677} \and Maria Fazio \inst{1}\orcidID{0000-0003-3574-1848} }

\authorrunning{R. Napoli}
%
\institute{University of Messina, Messina ME 98122, Italy 
}
\maketitle              
\begin{abstract}
Drones are embedded systems (ES) used across a wide range of fields, from photography to shipments and even during crisis management for searching, rescuing and damage assessment activities. However, their limited battery life and high energy consumption are very important challenges, especially in networked systems where multiple drones must communicate with a Ground Base Station (GBS). This study addresses these limitations by proposing the implementation of a bio-inspired leader-based energy management system for drone fleets. 
Inspired by bio-behavioral models, the algorithm dynamically chooses a single drone as a Leader in a cluster to handle long-range communication with the GBS, allowing other drones to preserve their energy.
The effectiveness of the proposed bio-inspired algorithm is evaluated by varying network sizes and configurations. The results demonstrate that our approach significantly increases network efficiency and service time by removing useless energy consumption communications. 

\end{abstract}
\section{Introduction}
Embedded Systems (ES) are electronic systems 
designed to meet a specific task.
They interact with the environment through sensors and actuators, satisfying strict requirements in terms of power consumption, size and cost.
Drones are small-medium size ES 
which enhance geolocalized data collection and computation. They are generally equipped with 
high-definition cameras and thermal sensors so that they can capture data even from areas difficult to access by humans~\cite{b1}, and are used in many different application fields, such as agriculture~\cite{b9}, emergency services~\cite{b10} and delivery systems~\cite{b11}. 

\begin{figure}[ht!]
    \centering
    \begin{subfigure}[b]{\textwidth}
        \includegraphics[width=\textwidth]{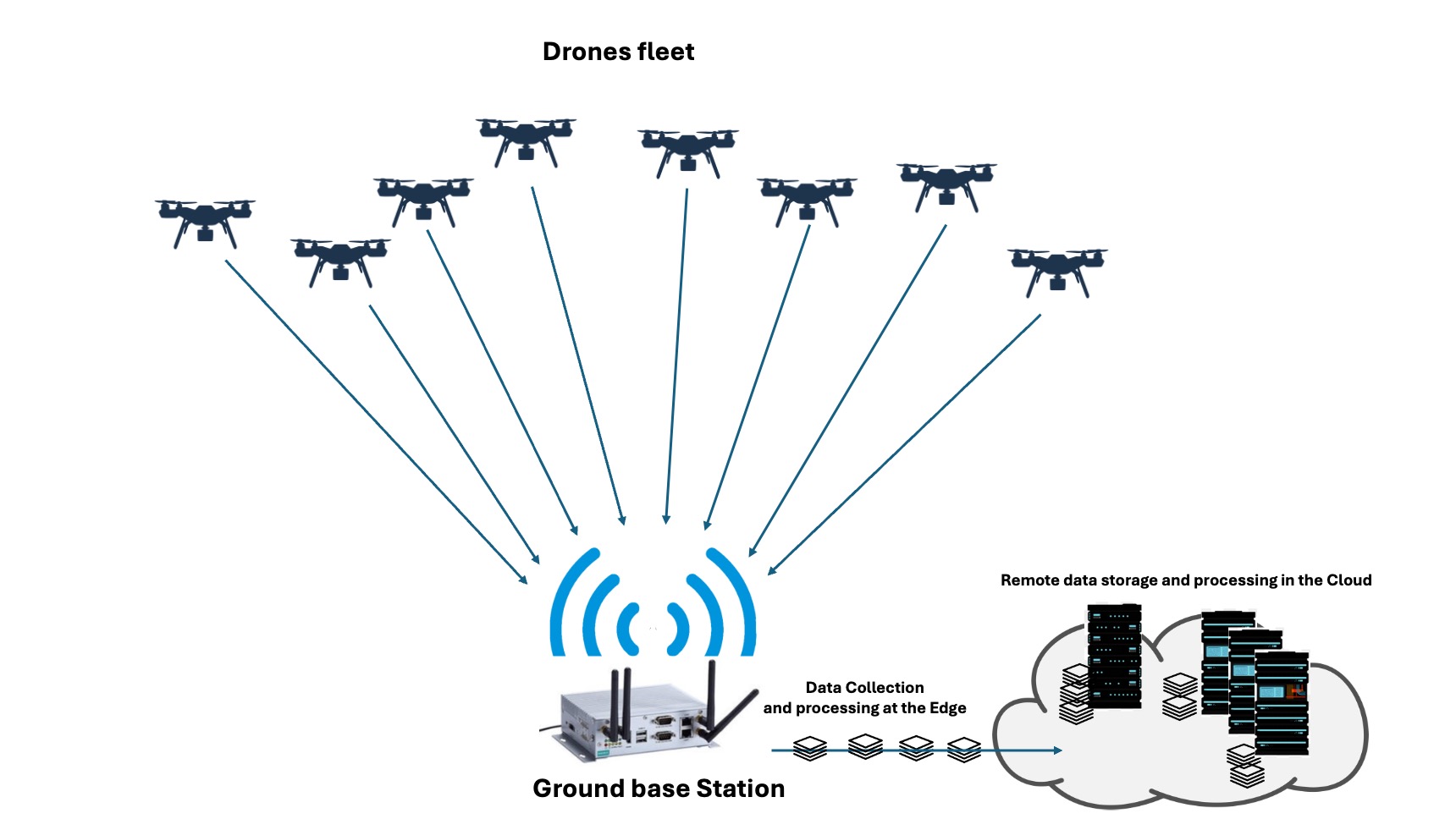}
        \caption{Typical Drones Network Architecture.}
        \label{fig:arch1}
    \end{subfigure}
\par\bigskip
\par\bigskip
    \begin{subfigure}[b]{\textwidth}
        \includegraphics[width=\textwidth]{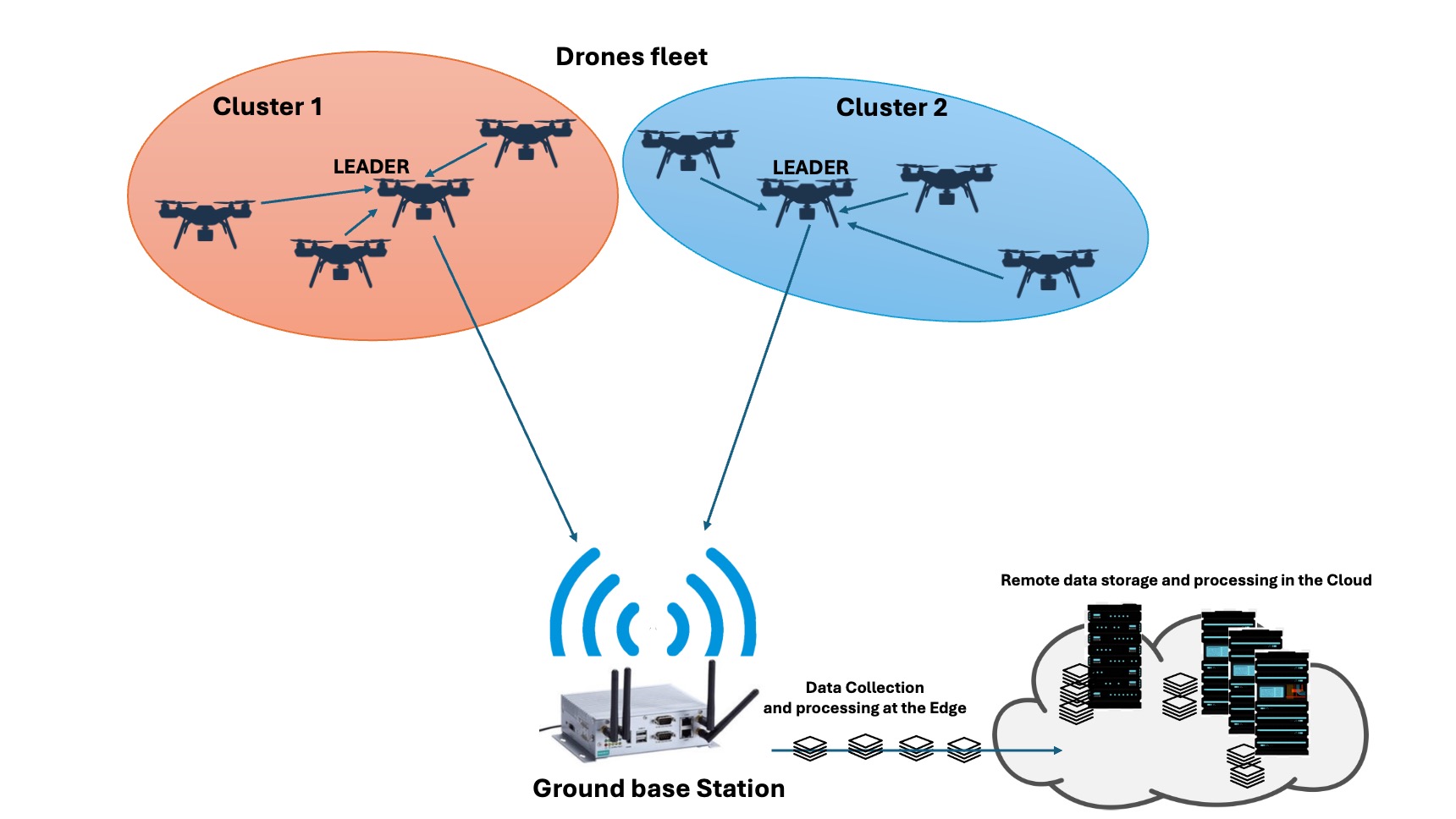}
        \caption{Leader-based Drones Network Architecture.}
        \label{fig:arch2}
    \end{subfigure}
    \caption{Different Architectures in Drone Fleet management.}
    \label{Typical Drones Network Architecture}
\end{figure}

In large-scale applications, drones typically work in groups, setting up a distributed computing environment, where they cooperate for a common goal, such as wildlife monitoring, search and rescue, firefighting and so on. To this aim, they communicate with each other to synchronize their activities, and with the Ground Base Station (GBS) for remote control and transfer of collected data. 
The GBS acts as a computing system at the Edge of the specific area where the drones are deployed and can interact with remote Cloud computing services for advanced processing
~\cite{b13} (see Figure~\ref{fig:arch1}).

Despite advancements in drone technologies, energy management is still a critical aspect that influences their efficiency and overall performance, so limiting the lifespan of the service they execute.
Energy Management Strategies (EMSs) identify methods to extend the life of a drone network by optimizing the average power consumption of the whole system. EMSs monitor consuming tasks and communication activities and implement different policies and tools for minimizing energy consumption~\cite{b5}.

In this paper, we present an EMS solution which aims to optimize the energy consumption of the drone network in the communication with the GBS. The basic idea starts from the consideration that generally, due to the altitude of drones, the long distance between the GBS and drones during flight becomes an issue for energy consumption, since transmission power in wireless communications increases with the square of the distance. 
Ground communication reduces the drone's battery life, causing poor energy performance. However, implementing a Leader-based communication of a cluster of close drones with the GBS can drastically reduce the energy consumption due to communication transmissions (see Figure~\ref{fig:arch2}). 


The solution we propose is based on a
Bio-inspired algorithms for Leader election using the principles of natural selection and evolution. 
This allows us to find optimal or near-optimal solutions to complex problems. 
By studying these natural phenomena, researchers have developed algorithms that mimic these behaviors to solve complex problems in more efficient ways, especially important in Internet of Things (IoT) applications and ES, where conditions can change rapidly and systems must adapt without human intervention, leading more resilient and efficient networks~\cite{b4}\cite{b16}. 
This paper explores the effectiveness of such bio-inspired EMS through simulations conducted on networks of varying sizes and battery thresholds, focusing on its ability to improve a network's lifespan and performance under different conditions.

\section{State of the Art}
In literature, some works deal with leader election in drone fleets to manage specific issues and, in particular, on how to improve the network's performance.
For example, in~\cite{b12}, a solution to connect drones to multiple GBSs maximizing communication throughput and minimizing energy consumption was proposed. To this aim, the optimal connection between drones and the nearest GBS is continuously calculated. 
In~\cite{b7}, the election is based on the Resource Strength Value (RSV), which serves as a capability measurement to determine the most suitable leader among drones in the fleets for EMS. The RSV was calculated using a matrix that incorporates various parameters reflecting the status of each drone, such as CPU and memory utilization.
However, the algorithm suits only homogeneous drone fleets, where drones have similar characteristics, as the matrix parameters for the RSV function have to be in the same ranges. 
The implementation of a proxy signature-based authentication mechanism in a drone fleet for device-to-device communication in 5G architectures is presented in~\cite{b17}. 
The leader is responsible for managing communications between the rest of the fleet and a 5G core network handling authentication, authorisation and access functions. 
Although it is an interesting work, implemented polices do not consider energy constrains. 
Trusted Dynamic Leader Selection (TDLS)~\cite{b18} is another algorithm based on a trusted drone in charge to manage communications in the fleet in a secure way.
The system uses a trust-based mechanism to identify trusted drones and isolate potentially malicious ones. 
However, from an EMS perspective, the exclusion of drones identified as malicious introduces a significant limitation to the system lifespan for energy constrains. 

Bio-inspired approaches are not yet widely used in literature but have already shown significant results. 
In~\cite{b19}, the way ants organise work within their colony was used to develop a distributed control approach, where control actions focus on the interactions between system components. 
The algorithm works around the Response Threshold Model (RTM), which is based on the way that ants react to external stimuli (such as the presence of food or the need to perform a task) according to a response threshold. 
In the control context, stimuli are represented by local variables, such as water levels in the tanks and the response threshold is a parameter that reflects the sensitivity of the controllers (or ants) to these stimuli. 
The experiments shows how the RTM-based control can adapt to different operational dynamics, showing better performance compared to classical control approaches. 
\cite{b20} presents a bio-inspired algorithm based on the Elephant Search Algorithm (ESA) to solve the NP-hard travelling salesman problem (TS) (i.e., given a list of cities and the distances between each pair of cities, the travelling salesman should find the shortest path to visits each city and return back to the first one). 
The proposed algorithm is based on observing the group dynamics of elephants, in particular how they form clans and follow a matriarch, who represents the most suitable elephant within the clan. 
The elephants within a clan update their positions according to the position of the matriarch, which is considered the current best fit. 
~\cite{b6} proposes a bio-inspired approaches for the coordination of drone fleets, achieving significant results in terms of energy savings (15\%\ more than classical approaches). 
Here, the leader not only has to manage communications with the GBS but also to coordinate and assign tasks to the other members of the fleet, resulting in a high communication overhead. Furthermore, a dynamic leader election mechanism based on adaptive thresholds is missing, as the algorithm 
is only activated when the leader's battery level falls below 40 per cent. 

\section{Bio-Inspired approach for Leader Election}
\label{sec:bioinspired}
The motivation behind this research is related to the potentiality of bio-inspired algorithms for making energy-efficient strategies in drone fleets, in order to create an efficient geo-localised data collection and processing system. As drones become integrated into everyday operations, the need for enhanced battery life has become crucial, so, giving bio-inspired distributed approaches to optimise the average life of the system, energy consumption per drone and poor network management can be overcome. 
To this aim, we have implemented a Bio-Inspired Leader-based EMS for Drone fleet management based on Particle Swarm Optimization (PSO). 

PSO is an algorithm that simulates the social behavior of birds searching for food, where individual agents (particles) adjust their position based on personal and general experiences so to preserve energy. This bio-inspired methodology not only enhances the search for optimal solutions but also increases adaptability in dynamic environments.
In PSO, each particle adjusts its position based on its own experience (the best solution it has found, known as $pBest$) and the experience of the entire swarm (the best solution found, known as $gBest$). Then, the particle with $gBest$ is chosen as the leader, taking on the responsibility for searching for food. This cooperation allows particles to share information and reach their task, avoiding useless waste of energy~\cite{b8}.
The core of the proposal is to choose a Leader within a cluster of drones that will manage the communication with the GBS, allowing the others to preserve energy for their main tasks. The algorithm selects a drone to act as the PSO $gBest$, relieving other drones from energy-intensive long-range communication, resulting in an extension of the overall network's lifespan. 

Considering a group of drones (the particles of PSO) with different battery levels, the goal is to have a $gBest$ drone that communicates with the GBS.
As for a swarm of birds, whose goal is to search for food, a fleet of drones has to communicate with the GBS. Having a leader who is in charge of managing the most energy-intensive task to protect the swarm, battery loss and the network's lifetime can be improved.
This model also ensures that no single drone is overburdened, as the role of the leader is rotated based on the battery level threshold. When the leader drone's battery level falls below a certain limit, a new leader is elected, allowing the former leader to preserve energy and extend the overall network’s life.
By conducting simulations with varying network sizes and configurations, this study aims to quantify the improvements in network lifespan and energy efficiency, giving the basis for further applications and contributing to significantly improve the efficiency of drone networks. 
As we will prove in this paper, by proposing this EMS within a cluster of drones, we can significantly reduce the overall energy consumption of the network, obtaining better results than other researches in this field.

\section{The proposed Energy Management Strategy}
Our work aims to improve the performance and lifespan of a fully connected drone network, also providing the effectiveness and convenience of using this EMS in this field. We assume that drones operate in networks and maintain a certain altitude for their tasks while keeping communication with the GBS.
However, as altitude increases, communication with their GBS becomes more energy-intensive as the wireless signal for communication must be amplified. Every drone in the network has to maintain direct communication with the GBS, causing significant energy consumption challenges for the entire network.
To address this, in our architecture (Fig \ref{fig:arch2}), the network is divided into $n$ clusters made of $m$ drones. A leader for each cluster is elected as described in Section~\ref{sec:bioinspired}, carrying out all the cluster's communications to the GBS. Positioned between the cluster and the GBS, the altitude required for communication is reduced, allowing the other non-leader drones to preserve their energy by avoiding direct ground communication.
All the messages from all the non-leader drones are stored inside a messages buffer. When the buffer is full or the leader must be changed, all the information is communicated to the GBS and then sent to a remote processing unit in the cloud.

Our model enhances efficiency by rotating the role of the leader drone. When the current leader’s battery level drops below a predefined threshold (based on the battery level), a new leader ($gBest$) is elected. This allows to the old leader to limit excessive battery consumption, distributing the power consumption among the cluster. By positioning the leader drone at an optimal altitude, halfway between the cluster and the GBS, the distance for communications to the ground is also reduced, limiting other unnecessary energy consumption.

\begin{figure}[ht!]
    \centering \includegraphics[width=0.9\linewidth]{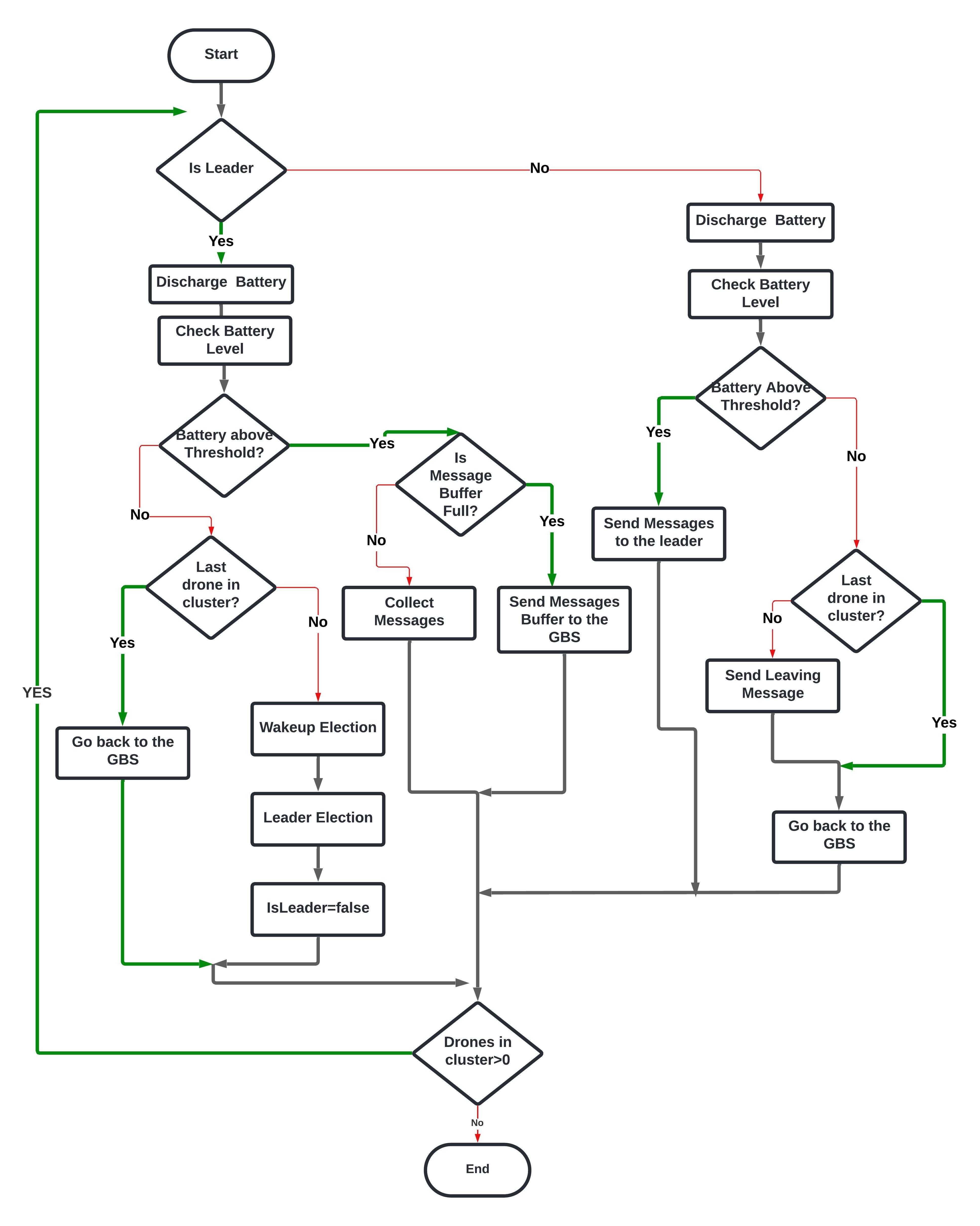}
    \caption{Drone's Routine}
    \label{fig:drones_routine}
\end{figure}

Compared to previous works in literature, there are no limitations regarding the homogeneity of drones in the network, as the $gBest$ parameter depends only on the discharge rate determined by the battery threshold, so this solution can be implemented on any kind of drone fleet. 
The algorithm designed for optimizing the battery life of drone networks is structured into several well-defined phases, each characterized by specific message types that facilitate communication and coordination among the drones.
Similar to how particles in PSO adjust their behaviour towards their personal best ($pBest$) and the global best ($gBest$), the drones monitor their battery levels and communicate with each other to identify the one with the highest charge. The leader would then act as the main node for communicating resources, giving all the fleet information to the GBS (as in PSO for food source).
Both systems rely on the exchange of information to optimize their respective objectives. PSO seeks to find the optimal solution for finding food and preserving energy, while the proposed algorithm aims to designate a leader that can efficiently manage messages exchanged with the GBS. This comparison highlights the adaptive nature of both methodologies in responding to dynamic environments and optimizing performance through collaborative behavior.
Each drone has a routine that dynamically changes according to the role it plays (see Figure \ref{fig:drones_routine}).

The first leaders of the simulation are designated as the drones with the lowest identifiers (IDs), which is a common practice in this kind of algorithms. In particular, the leader election is based on a threshold value, which defines the limit of the battery in order to change the current leader.
Formally, we introduce the threshold as a percentage value that defines the limit at which a new leader is elected. Let $B_0$ be the battery capacity of a drone when it is elected as leader, the new election will start when its battery level drops below the threshold value $B_T$.
$B_T$ is given by the following equation:
\begin{equation}
    B_T = B_0 \times \left( 1 - \frac{T}{100} \right)
    \label{eq:BT_equation}
\end{equation}

Where:
\begin{itemize}
    \item $B_0$ is the drone's battery capacity when it becomes leader, expressed in milliampere-hours (mAh).
    \item $T$ is the threshold value, expressed as a percentage (e.g., \( T = 60\%\)).
\end{itemize}
Therefore, once the battery reaches $B_T$, a new leader is elected. 
The algorithm is divided into 5 main phases: 1)~\textit{Initialization}, 2)~\textit{Update}, 3)~\textit{Election Initialization}, 4) \textit{Election} and 5) \textit{Termination}. 

\begin{algorithm}[ht!]
\label{algorithm:init_phase}
\SetAlgoLined
\KwResult{Drones initialization}
\textbf{Initialization Phase:}
\If{Drone receives an initMessage}{
    \If{Drone is the Leader}{
        Set the leader's position\;
        Communicates its role to all the other drones within the same cluster\;
    }
    \Else{
        \If{A leader message has arrived}{
            Start the update phase\;
        }
        \Else{
        Wait a message from the leader\;
        }

    }
}
\caption{Bio-Inspired Leader-based EMS Algorithm's Initialization Phase}
\end{algorithm}

\vspace{10pt}

\paragraph{\textbf{Initialization Phase}}
Leaders broadcast an initialization message to the other drones within their respective clusters. This message prompts the drones to begin their routines (see Algorithm 1).

\vspace{10pt}

\begin{algorithm}[H]
\label{algorithm:update_phase}
\SetAlgoLined
\KwResult{Update Phase}
\textbf{Update Phase:}
\If{Drone is the leader}{
    \If{Messages Buffer is full}{
        Send the Messages Buffer collected information to the GBS\;
        Reset the Messages buffer\;
    }\Else{
        Collect data\;
    }
}
\Else{
    Send information to the leader\;
}
\caption{Bio-Inspired Leader-based EMS Algorithm's Update Phase}
\end{algorithm}

\vspace{10pt}

\paragraph{\textbf{Update Phase}}
Non-leader drones continuously send data in the form of messages to their leader, who communicates with the GBS.
The leader has a buffer that collects data from the other drones and when it is full, it sends a "baseMessage" to the GBS. The remaining battery levels of the leaders are monitored to determine when a new election is needed, by analyzing the threshold value with respect to the current battery level (see Algorithm \ref{algorithm:update_phase}). 

\vspace{10pt}

\begin{algorithm}[H]
\label{algorithm:election_initialization_phase}
\SetAlgoLined
\KwResult{Managing of Election initialization Phase }
\textbf{Election Initialization Phase:}
\If{Battery level is below a certain threshold OR Leader triggers an election}{
    \If{Drone is Leader}{
        Send WakeupElectionMessage to all drones in the cluster\;
        Wait for responses from other drones\;
        Compare battery levels and select the new Leader\;
        Start election phase\;
    }
    \Else{
        Send battery and position information to the leader\;
    }
}
\caption{Bio-Inspired Leader-based EMS Algorithm's Election initialization Phase.}
\end{algorithm}

\vspace{10pt}

\paragraph{\textbf{Election initialization Phase}}
When a leader's remaining battery level reaches the threshold (see Equation \ref{eq:BT_equation}), it sends its buffer to the GBS an then begins the election initialization phase. A "wakeupElection" message is sent and each drone within the cluster replays with its: ID, cluster and remaining battery level. The leader evaluates these responses to determine who is the $gBest$, then it starts the election phase (see Algorithm \ref{algorithm:election_initialization_phase}).

\vspace{10pt}

\begin{algorithm}[ht!]
\label{algorithm:election_phase}
\SetAlgoLined
\KwResult{Complete drone and algorithm behavior}
\textbf{Election Phase:}
\If{Drone receives ElectionMessage}{
    \If{Drone ID matches the new Leader ID}{
        Assume the Leader role\;
        Update position and battery threshold\;
    }
    \Else{
        Update my new Leader and start Update phase\;
    
    }
}
\caption{Bio-Inspired Leader-based EMS Algorithm's Election Phase Phase.}
\end{algorithm}

\vspace{10pt}

\paragraph{\textbf{Election Phase}}
The current leader broadcasts an election message to all nodes within the cluster, announcing the new $gBest$. After receiving this message, each drone updates its global knowledge. 
A new update phase starts and when the Equation \ref{eq:BT_equation} is satisfied by the new leader, it eventually triggers a new Election Initialization Phase by sending a "wakeUpElection" message to all nodes (see Algorithm \ref{algorithm:election_initialization_phase}). 

\vspace{10pt}

\paragraph{\textbf{Termination Phase}}
The algorithm includes a termination phase that is triggered when a drone’s battery reaches a critically low level, indicating that it must return to its base station for recharging. When a non-leader drone needs to leave the cluster, it sends a "leaveMessage" to the current leader. After receiving this message, the leader updates its local knowledge, reducing the cluster size by one.  Instead, if the leader has to leave and the cluster size is greater than one, a new election starts, otherwise the algorithm ends.

\vspace{10pt}

The algorithm's design emphasizes energy efficiency by dynamically electing leaders based on battery levels, thereby optimizing communication and extending the overall lifespan of each cluster. This structured approach ensures a balanced distribution of energy consumption across the network, leading to improved performance and sustainability in various applications.

\section{Results}
OMNeT++ is a message-based discrete event simulation environment and one of the most famous open-source simulators in the computer science field~\cite{b14}.
In particular, among OMNeT++ v.6.0.3, the INET framework was used to make simulations, as it offers a wide set of models for simulating different types of communication networks, through the possibility of managing in an easy way nodes and edges~\cite{b14}.

The network was modelled by defining a simple Omnet++ module called $Drone$ that represents each drone within the simulation. This module includes parameters for managing: position, local battery, leader battery and leader-id. Each drone also has attributes to monitor and handle communication, like message rate and cluster size.
The topology network was made by modifying the INet NED file and it is characterized by $n$ cluster containing $m$ drones (based on the experiment).  All drones within a cluster are interconnected, allowing direct communication between every pair of drones. This connection was modelled using the NED file for managing edges and creating bidirectional links between drones.

\begin{algorithm}[ht!]
\SetAlgoLined
    \KwResult{Network Topology in Omnet++ simulations}    
    \Begin{
        \For{each cluster $c_i$ in network}{
            \For{each drone $d_i$ in cluster $c_i$}{
                Initialize drone $d_i$ with cluster\_id $c_i$\;
                Connect drone $d_i$ to other drones in cluster\;
            }
        }
        
        \For{each drone pair $(d_i, d_j)$ where $i \neq j$ and $j > i$}{
            Connect $d_i.gate++$ with $d_j.gate++$\;
        }
    }
    
    \caption{NED Network Topology}
\end{algorithm}

The results of the simulations conducted to evaluate the effectiveness of the bio-inspired leader-based energy management system are presented in this section. The analysis focuses on various scenarios, including the impact of different battery thresholds for leader election and the number of drones in the network.

The first experiment focuses on the average cluster lifetime in terms of the Omnet++ time step as the number of drones in the network increases. The experiment aims to compare the results obtained with classical architecture, in which each drone communicates with the ground base, against the proposed innovative model with different threshold values.

The parameter for electing a new leader (as in the case of PSO's $gBest$) is the threshold value. It indicates when the drone has used a lot of energy and must be changed to avoid overloading it too much, causing it to discharge too quickly. Several experiments have been conducted on this value in order to understand how it actually affects communications management and network lifetime.

\begin{figure}[ht!]
    \centering
    \includegraphics[width=0.9\linewidth]{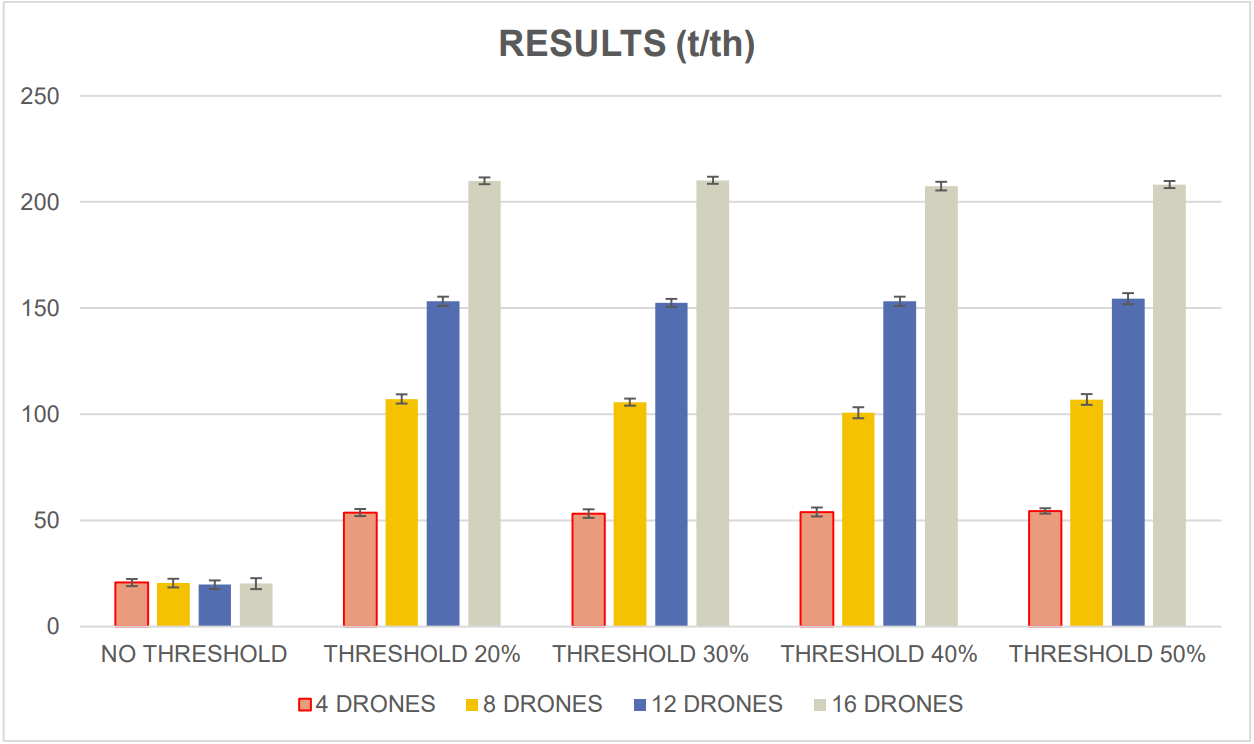}
    \caption{Time/Threshold Results.}
    \label{Time/Threshold Results}
\end{figure}

The results in Figure \ref{Time/Threshold Results} show how the algorithm actually contributes significantly to increasing the average network lifetime.
This suggests that the energy management system works as expected as the communication with the GBS is distributed among the different leaders during the activity, making it a promising approach for applications requiring continuous service.

\begin{figure}[ht!]
    \centering
    \includegraphics[width=0.7\linewidth]{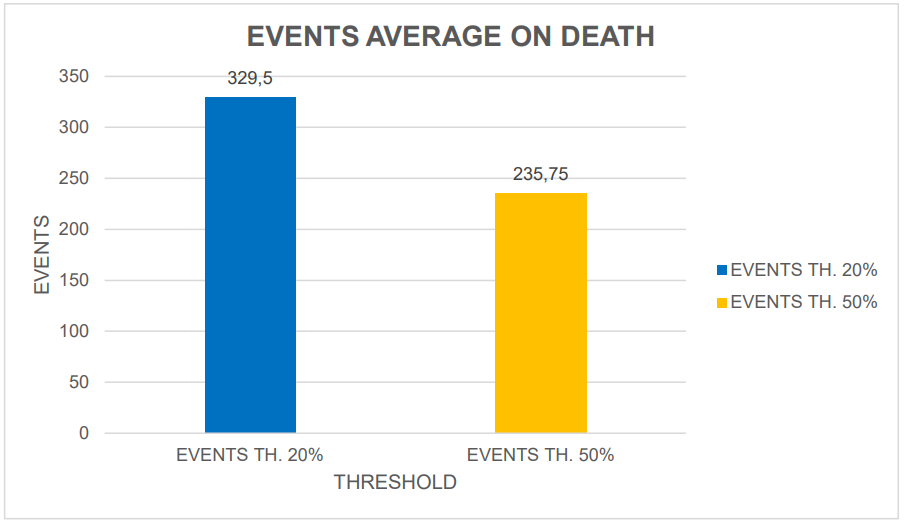}
    \caption{Event on death}
    \label{events_on_death}
\end{figure}

The threshold values analysis made on events on death (see Figure \ref{events_on_death}) revealed that reducing the threshold did not significantly modify the lifespan of each drone but it led to a higher number of messages exchanged, due to more frequent elections in the end of network's lifespan. This means that a balance between the number of election messages and the operational ones is needed in order to avoid messages overhead. In further applications, this overhead can be used to piggybacking other useful messages but it depends on the application. 

\begin{figure}[ht!]
    \centering
    \includegraphics[width=0.7\linewidth]{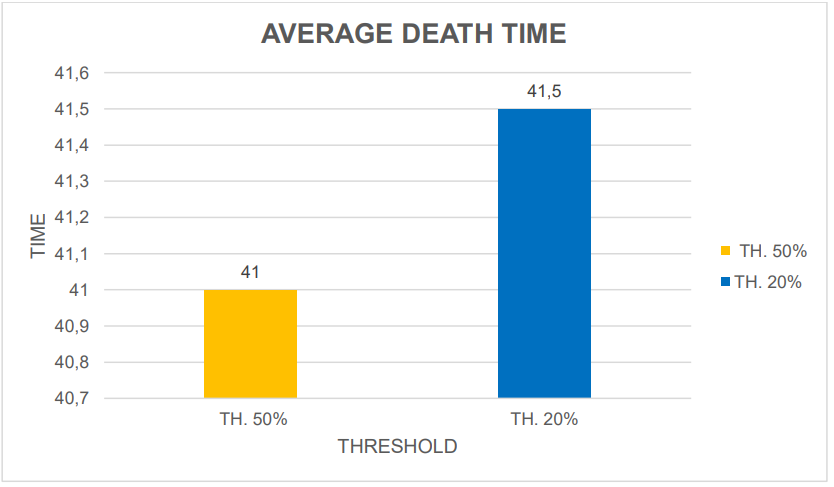}
    \caption{Average Death Time}
    \label{fig:Average_Death_Time}
\end{figure}

In another experiment, we focused on the average lifetime of each drone at varying thresholds and the average death time did not vary a lot across different threshold settings (see Figure \ref{fig:Average_Death_Time}). This means that lower thresholds resulted in a higher election frequency and messages complexity within the network, not changing the average lifetime of the cluster, as election messages are not very energetic because they are exchanged between leader and non-leader drones.

\begin{figure}[ht!]
    \centering
    \includegraphics[width=0.8\linewidth]{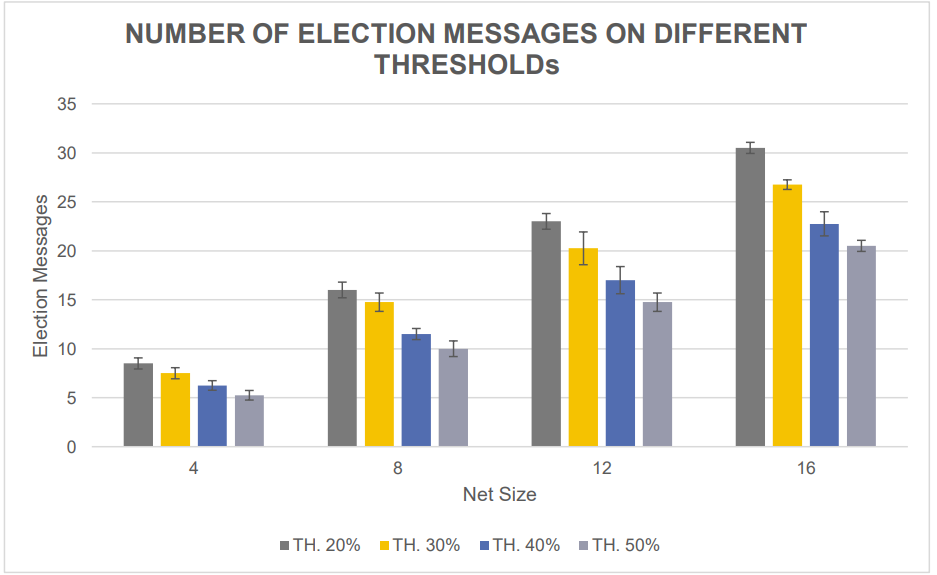}
    \caption{Number of election messages on different thresholds}
    \label{fig:election_messages_on_different_th}
\end{figure}

A final analysis was conducted to evaluate the increase in the number of election messages as the threshold changes (see Figure \ref{fig:election_messages_on_different_th}). As the threshold decreases, the number of elections increases because the leader switch condition, in the final stages of the cluster's life, is reached faster.
These results suggest that low level of thresholds cause frequent elections that enhance responsiveness and communication overhead. The balance has to be found with respect to the final task. For example, in crisis management more network responsiveness is required in face off message complexity overhead. 

\section{Conclusions}
The research presented in this paper demonstrates the effectiveness of our leader-based energy management system in enhancing the overall lifespan of drone networks. Through a series of simulations, it has been established that the implementation of this algorithm significantly increases the efficiency and duration of the network, thereby addressing one of the primary limitations associated with drone battery life.
In addition, as the number of drones increases, the average duration of the cluster also increases, suggesting that the distributed battery system allows better management of communications and, therefore, of each individual drone's battery.

The results are particularly promising, as they suggest that using more drones can lead to greater efficiency and overall autonomy of the system.
By reducing the threshold value, the individual lifespan of each individual drone does not change but the number of messages exchanged is higher because of the increasing of messages complexity. This means that we have to pay attention to the main goal of our service, in order to have a good balance between the number of elections and messages exchanged.
This can be advantageous in some specific applications, such as image capture, where a drone cluster must be active for as long as possible, as well as optimize its period of activity.
In the end, threshold values also represent a way to prevent drones from completely draining the battery in extreme situations. 

In real applications, it is necessary for drones to re-enter the base for charging, so having also a minimum battery limit threshold allows them to leave the network in safety.
In conclusion, the implementation of the bio-inspired leader-based energy management system demonstrated a significant increase in overall network life and can be the first step for further works to achieve greater energy efficiency and optimized lifespan of drone networks.
%
%
%

\begin{thebibliography}{8}
\bibitem{b1} C. Kyrkou, S. Timotheou, P. Kolios, T. Theocharides and C. Panayiotou, "Drones: Augmenting Our Quality of Life," in IEEE Potentials, vol. 38, no. 1, pp. 30-36, Jan.-Feb. 2019, doi: 10.1109/MPOT.2018.2850386.
keywords: {Drones;Payloads;Cameras;Monitoring;Thermal sensors;Laser radar},
\bibitem{b2} Mohd Daud, S. M. S., Mohd Yusof, M. Y. P., Heo, C. C., Khoo, L. S., Chainchel Singh, M. K., Mahmood, M. S.,  Nawawi, H. (2022). Applications of drone in disaster management: A scoping review. Science and Justice, 62(1), 30-42. https://doi.org/10.1016/j.scijus.2021.11.002,
\bibitem{b3} Al-Naji, A.; Perera, A.G.; Mohammed, S.L.; Chahl, J. Life Signs Detector Using a Drone in Disaster Zones. Remote Sens. 2019, 11, 2441. https://doi.org/10.3390/rs11202441

\bibitem{b4} Hamidouche, R., Aliouat, Z., Gueroui, A. M., Ari, A. A. A., and Louail, L. (2018). Classical and bio-inspired mobility in sensor networks for IoT applications. Journal of Network and Computer Applications, 121, 70–88. https://doi.org/10.1016/j.jnca.2018.07.010

\bibitem{b5} Tlili, F., Fourati, L. C., Ayed, S.,  Ouni, B. (2022). Investigation on vulnerabilities, threats and attacks prohibiting UAVs charging and depleting UAVs batteries: Assessments and countermeasures. Ad Hoc Networks, 129, 102805. https://doi.org/10.1016/j.adhoc.2022.102805

\bibitem{b6} Ganesan, R.; Raajini, X.M.; Nayyar, A.; Sanjeevikumar, P.; Hossain, E.; Ertas, A.H. BOLD: Bio-Inspired Optimized Leader Election for Multiple Drones. Sensors 2020, 20, 3134. https://doi.org/10.3390/s20113134

\bibitem{b7} C. Wanniarachchi, P. Wimalaratne and K. Karunanayaka, "A Behavioral Model for Single Leader: Leader-Follower Drone Swarm in Leader’s Failure," 2023 IEEE 2nd Industrial Electronics Society Annual On-Line Conference (ONCON), SC, USA, 2023, pp. 1-6, doi: 10.1109/ONCON60463.2023.10431018.

\bibitem{b8}R. Eberhart and J. Kennedy, "A new optimizer using particle swarm theory," MHS'95. Proceedings of the Sixth International Symposium on Micro Machine and Human Science, Nagoya, Japan, 1995, pp. 39-43, doi: 10.1109/MHS.1995.494215.

\bibitem{b9} Van der Merwe, Deon and Burchfield, David and Witt, Trevor and Price, Kevin and Sharda, Ajay. (2020). Drones in agriculture. 10.1016/bs.agron.2020.03.001. 

\bibitem{b10} Wankmüller, Christian  Kunovjanek, Maximilian  Mayrgündter, Sebastian. (2021). Drones in emergency response – evidence from cross-border, multi-disciplinary usability tests. International Journal of Disaster Risk Reduction. 65. 102567. 10.1016/j.ijdrr.2021.102567. 

\bibitem{b11} Luigi Di Puglia Pugliese, Luigi Guerriero, Francesca Macrina, Giusy. (2020). Using drones for parcels delivery process. Procedia Manufacturing. 42. 488-497. 10.1016/j.promfg.2020.02.043. 

\bibitem{b12} Xu, Yu, Xiao, Lin, Yang, Dingcheng, Cuthbert, Laurie, Wang, Yapeng, Energy-Efficient UAV Communication with Multiple GTs Based on Trajectory Optimization, Mobile Information Systems, 2018, 5629573, 10 pages, 2018. https://doi.org/10.1155/2018/5629573

\bibitem{b13} Wu, Huifeng, Hu, Junjie, Sun, Jiexiang, Sun, Danfeng, Edge Computing in an IoT Base Station System: Reprogramming and Real-Time Tasks, Complexity, 2019, 4027638, 10 pages, 2019. https://doi.org/10.1155/2019/4027638

\bibitem{b14} Bachmeier, S., Jaeger, B., Holzinger, K. (n.d.). Network Simulation with OMNet++. https://doi.org/10.2313/NET-2020-11-08

\bibitem{b15} De Micco, L., Vargas, F. L.,  Fierens, P. I. (2020). A Literature Review on Embedded Systems. In IEEE Latin America Transactions (Vol. 18, Issue 2, pp. 188–205). IEEE Computer Society. https://doi.org/10.1109/TLA.2020.9085271


\bibitem{b16}
G. Devika and A. G. Karegowda. Bio-inspired optimization: algorithm, analysis and scope of application. IntechOpen, Rijeka, 2023.

\bibitem{b17} Abdel-Malek, M. A., Akkaya, K., Bhuyan, A.,  Ibrahim, A. S. (2022). A Proxy Signature-Based Swarm Drone Authentication With Leader Selection in 5G Networks. IEEE Access, 10, 57485–57498. https://doi.org/10.1109/ACCESS.2022.3178121

\bibitem{b18} Kundu, J., Alam, S., Koner, C., Piran, M. J. (2024). Trust-Based Dynamic Leader Selection Mechanism for Enhanced Performance in Flying Ad-Hoc Networks (FANETs). IEEE Transactions on Intelligent Transportation Systems. https://doi.org/10.1109/TITS.2024.3456132

\bibitem{b19} García, M., Aguilar, J. (2023). A bio-inspired emergent control approach for distributed processes. Applied Soft Computing, 141.https://doi.org/10.1016/j.asoc.2023.110318

\bibitem{b20} Hossam, A., Bouzidi, A., Riffi, M. E. (2019). Elephants herding optimization for solving the travelling salesman problem. Advances in Intelligent Systems and Computing, 912, 122–130. https://doi.org/10.1007/978-3-030-12065-8-121227-016-1739-2.Springer, New York.
\end{thebibliography}
%

\end{document}